\documentclass{aastex}
\usepackage{spr-astr-addons}
\usepackage{graphicx,epsfig}

\begin{document}

\title{A Universal Vertical Stellar Density Distribution Law for the Galaxy}
\shorttitle{A Single Density Law}
\shortauthors{S. Karaali, E. Hamzao\u glu, S. Bilir}

\author{S. Karaali \altaffilmark{1}}
\altaffiltext{1}{Beykent University, Faculty of Science and Letters, Department 
of Mathematics and Computer, Beykent Ayaza\u ga Campus, 34398, Istanbul, Turkey\\
\email{skaraali@beykent.edu.tr}}
\and
\author{E. Hamzao\u glu\altaffilmark{2}} 
\altaffiltext{2}{Beykent University, Department of Management Information Systems, 
Beykent Ayaza\u ga Campus, 34398, Istanbul, Turkey\\}

\and
\author{S. Bilir\altaffilmark{3}}
\altaffiltext{3}{Istanbul University, Faculty of Science, Department 
of Astronomy and Space Sciences, 34119 University, Istanbul, Turkey}


\begin{abstract}
We reduced the observational logarithmic space densities in the vertical 
direction up to 8 kpc from the galactic plane, for stars with absolute 
magnitudes (5,6], (6,7] and [5,10] in the fields $\#$0952+5245 and SA114, to a single 
exponential density law. One of three parameters in the quadratic expression 
of the density law corresponds to the local space density for stars with 
absolute magnitudes in question. There is no need of any definition for 
scaleheights or population types. We confirm with the arguments of non–discrete 
thin and thick discs for our Galaxy and propose a single structure up to 
several kiloparsecs from the galactic plane. The logarithmic space 
densities evaluated by this law for the ELAIS field fit to the observational 
ones. Whereas, there are considerable offsets for the logarithmic space 
densities produced by two sets of classical galactic model parameters from 
the observational ones, for the same field.

\end{abstract}


\section{Introduction}
For some years, a disagreement exists among the researchers about the 
formation history of our Galaxy. Yet there has been a large improvement 
about this topic since the pioneering work of \citet{ELS62} who argued that 
the Galaxy collapsed in a free-fall time ($\sim 2\times10^{8}$ yr). 
We know that the Galaxy collapsed over many Gyr \citep[e.g.][]{Y79, 
N85, N86, S87, C90, N91, B95} and at least some of the components are formed 
from the merger or accretion of numerous fragments, such as dwarf type galaxies 
\citep[cf.][and references therein]{S87, F02}. Also, the number of population 
components of the Galaxy increased by one, complicating interpretations of any 
data set. The new component (the thick disc) was introduced by \citet{G83} in 
order to explain the observations that star counts towards the south galactic pole 
were not in agreement with a single-disc (thin-disc) component, but rather could 
be much better represented by two such components. This was the simplest 
combination of free parameters giving a satisfactory fit, and simplicity  
can play a major role in astrophysical fits. Different parameterization followed 
the work of \citet{G83}. For example, \citet{K89} showed that the vertical structure 
of our Galaxy could be best explained  by a multitude of quasi-isothermal components, 
i.e. a large number of sech$^{2}$ isothermal discs, together making up a more 
sharply peaked sech or exponential distribution. Also, we quote the work of 
\citet{deGrijs97} who made clear that in order to build up a sech distribution, 
one needs multiple components. Finally, we quote our work \citep{K06} where the 
galactic structure were parameterized by two exponentials up to $z\sim10$ kpc, 
covering thin disc, thick disc and inner halo. 

Although different parameterization were tried by many researchers, only the one 
of \citet{G85} which is based on star counts estimation for thin disc, thick disc 
and spheroid (halo) became as a common model for our Galaxy, and used widespread 
with improving the parameters, however. In other words, the canonical density 
laws are as follows: 1) a parameterization for thin and thick discs in cylindrical 
coordinates by radial and vertical exponentials and 2) a parameterization for halo 
by the \citet{deVaucouleurs48} spheroid. The thin disc dominates the small $z$ 
distances from the galactic plane with a small scaleheight, whereas the thick 
disc extends to larger $z$ distances with larger scaleheight. In some studies, 
the range of values for the parameters is large, especially for the thick disc. 
For example, \citet{C01} and \citet{S02} give 6.5-13 and 6-10 per cent, 
respectively, for the relative local density for the thick disc. In the paper 
of \citet{KBH04}, we discussed the large range of these parameters and claimed 
that galactic model parameters are absolute magnitude dependent. We showed that 
the range of the model parameters estimated for a unique absolute magnitude 
interval is considerably smaller. 

It is true that the kinematical and metallicity structure of two regions in our 
Galaxy is different. Particularly, high metallicity stars lie relatively closer 
to the galactic plane than with low metallicity stars. Also, kinematical 
dispersions of these stars are smaller than the stars which occupy the distant 
regions of the Galaxy. Does that mean, above described differences suggest the 
separation of the Galaxy into different populations such as thin and thick discs?  
\citet{N87} proposed a Galaxy model which does not assume that thin and thick discs 
are discrete components, but instead form a kinematical and chemical continuum. 
Stars traditionally associated with the thick disc belong to an ``extended'' disc 
(in terms of spatial distribution)in Norris' terminology, and represent an extreme 
tail of metallicity and kinematic distribution. In the work of \citet{I08}, the 
absence of a correlation between the observed velocity and metallicity distribution 
for disc stars represents a major problem for the interpretation of vertical 
velocity and metallicity gradients as being due to a varying linear combination 
of two fixed distributions. These authors argue that their results appear roughly 
consistent with Norris' proposal. The works of \citet{N87} and \citet{I08} are 
two examples, where discreteness of the thin and thick discs are discussed.

The studies cited in the previous paragraph encouraged us to propose a density 
law with three parameters without separating the Galaxy into different components. 
The procedure followed in this work is different than the previous study presented 
\citep{K06}. Here, for the sake of simplicity, only vertical space densities were 
used. However, we shall cover radial density variations in near future.   

In Sections 2 and 3, the canonical density law forms and the new density law are 
discussed. The calibration of the newly defined galactic model parameters for 
the new density law is given in Section 4. In Section 5, the new density law is 
tested on the space densities of a different field. Section 6 provides a discussion 
and finally a conclusion is given in Section 7.   

\section{The canonical density law forms}
Disc structures are usually parameterized in cylindrical coordinates by radial and 
vertical exponentials,
\begin{eqnarray}
\tiny
D_{i}(x,z)=n_{i}\exp(-z/H_{i})\exp[-(x-R_{o})/h_{i}],
\end{eqnarray}
where $z$ is the distance from galactic plane, $x$ is the planar distance 
from the galactic center, $R_{0}$ is the solar distance to the galactic 
center \citep[8 kpc,][]{R93}, $H_{i}$ and $h_{i}$ are the scaleheight and scalelength 
respectively, and $n_{i}$ is the normalized local space density. The suffix $i$ 
takes the values 1 and 2, as long as thin and thick discs are considered. 
A similar form uses the sech$^{2}$ (or sech) function to parameterize 
the vertical distribution for thin disc,

\begin{eqnarray}
D_{i}(x,z)=n_{i}{\rm sech}^{2}(z/H^{'}_{i})\exp[{-(x-R_{o})/h_{i}}].
\end{eqnarray}
Because the sech function is the sum of two exponentials, $H^{'}_{i}$ is 
not really a scaleheight, but, it has to be compared to $H_{i}$ by multiplying 
it with 2: $H_{1}=H^{'}_{1}/2$ \citep{vanderkruit81a, vanderkruit81b, vanderkruit82a,
vanderkruit82b, vanderkruit88}. However, in order to build up such a distribution, 
one needs multiple components \citep{K89, deGrijs97}.

The density law form for the spheroid (halo) component is parameterized in different 
forms. The most common is the \citet{deVaucouleurs48} spheroid used to describe the 
surface brightness profile of elliptical galaxies. This law has been de–projected 
into three dimensions by \citet{Y76} as 
\begin{eqnarray}
D_{s}(R)=n_{s}\exp[-7.669(R/R_{e})^{1/4}]/(R/R_{e})^{7/8},
\end{eqnarray}
where $R$ is the (uncorrected) galactocentric distance in spherical coordinates, 
$R_{e}$ is the effective radius and $n_{s}$ is the normalized local space density. 
$R$ has to be corrected for axial ratio $\kappa=c/a$, 
\begin{eqnarray}
R = [x^{2}+(z/\kappa)^2]^{1/2},
\end{eqnarray}
where,
\begin{eqnarray}
x = [R_{o}^{2}+(z/\tan b)^2-2R_{o}(z/\tan b)\cos l]^{1/2},
\end{eqnarray}
$b$ and $l$  being the galactic latitude and longitude, respectively, for the 
field under investigation.

An alternative formulation is the power law,
\begin{eqnarray}
D_{s}(R)=n_{s}/(a_{o}^{n}+R^{n}),
\end{eqnarray} 
where $a_{0}$ is the core radius.

If one restricts the work to the vertical direction, then the third factor in 
Eqs. (1) and (2) can be neglected.

\section{A unified density law for the thin disc, thick disc and halo}	 	
As already mentioned in Section 1, the introduction of the thick disc component into 
the studies was in order to better representation of the observed star counts in the 
south galactic pole. However, the metallicities and kinematical data of thin and 
thick discs overlap. This is contradictory to the arguments related to two discrete 
components. What we mean, when we restrict our work to vertical direction by canonical 
approach, is the space density which decreases by amount equivalent to $\exp (-1)$ 
at each distance adopted for the thin disc up to a $z$-–distance from the 
galactic plane where the thin disc is dominant. Whereas where thick disc dominates, 
the space density decreases with smaller gradient at larger $z$--distances.   

Thus, two components such as thin and thick discs were arbitrarily adopted just to fit 
the observational data, up to a few kpc. Then, we can approach in a similar way, 
i.e. we can introduce a density law which fits to the observational data up to a 
distance from the galactic plane, without defining any population type, however. Also, 
there are some indications that, a significant fraction of material with $[Fe/H]<-1$ dex 
has disc like structure \citep{N86}. Hence, we expect a larger $z$--distance interval for 
the new density law we propose in the following.   

Our new density law covers the canonical density law, for densities in the 
vertical direction and it contains no constant scaleheight. In the vertical 
direction, the density law form for thin and thick discs takes the form as,

\begin{eqnarray}
D(z)=n \exp(-z/H).
\end{eqnarray}
which can also be written as,

\begin{eqnarray}
D(z)=\exp[-(a_{1}z + a_{0})],
\end{eqnarray}
where $a_{1}=1/H$ and $n=\exp(-a_{0})$. The parameter $a_{0}$ is constant due to 
its definition. However, we let $a_{1}$ which is defined as the reciprocal of the 
scaleheight $H$ in situ, to change with $z$–-distance. Hence, we assume that the 
scaleheight changes continuously rather than as a step function. This is the main 
difference between the arguments of our work and the ones in situ.

When we examine Eq. (8), we see that the space density in vertical direction 
changes exponentially as a linear function of $z$. We consider a quadratic function 
of $z$ could be better matched to observational data. Hence, we obtained the final 
density law by adding a quadratic term to Eq. (8):

\begin{eqnarray}
D(z)=\exp[-(a_{2}z^{2} + a_{1}z + a_{0})].
\end{eqnarray}
For the estimation of the coefficients $a_{i}$ ($i$=0,1,2), the following 
procedure was adopted. Eq. (9) is written in logarithmic form:

\begin{eqnarray}
a_{2}z^{2} + a_{1}z + a_{0}=-\ln D(z).
\end{eqnarray}
Here, we used Hipparcos' local space density \citep{J97} for a specific absolute 
magnitude interval $M_{1}–-M_{2}$ and a sequence of observed $D(z)$ space 
densities and estimated the corresponding coefficients.

\section{Estimation of the parameters in the new density law}
	
The Eq. (10) applied to the data taken from \citet{KBH04}. The first set were 
evaluated for stars with absolute magnitude intervals (5,6] and (6,7], for the 
field $\#$0952+5245 (Table 1), whereas the second set covers the space densities 
for a unique absolute magnitude interval for stars with absolute magnitudes (5,10] 
for the field SA 114 (Table 2). Here, $D^{*}=\log D+10$, $D=N/\Delta V_{1,2}$; 
$\Delta V_{1,2}=(\pi/180)^{2}(A/3)(r_{2}^{3}-r_{1}^{3})$; A denotes the size 
of the field; $r_{1}$ and $r_{2}$ denote the limiting distance 
of the volume $\Delta V_{1,2}$; $N$ denotes the number of stars in this volume; 
and $z^{*}=r^{*}\sin(b)$, is the vertical distance of the centroid of the volume 
$\Delta V_{1,2}$ where $b$ is the galactic latitude of the field and $r^{*}$ the 
distance of the centroid in the line of sight. The logarithmic space densities 
$D^{*}=7.47$ in Table 1 and $D^{*}=7.52$ in Table 2 were evaluated by the Hipparcos' 
local space density \citep{J97}.

\begin{table}
\center
\caption{Distance to the galactic plane ($z^{*}$ in kpc) and the logarithmic 
space density ($D^{*}$) data for two absolute magnitude intervals for the 
field $\#$0952+5245 ($l=83.^{o}38$, $b=48.^{o}55$).}
\begin{tabular}{ccccc}
\hline
$M_{g}\rightarrow$ & \multicolumn{2}{c}{(5,6]} &   \multicolumn{2}{c}{(6,7]}\\
\hline
        ID &    $z^{*}$ &    $D^{*}$ &    $z^{*}$ &    $D^{*}$ \\
\hline
         1 &       0.00 &       7.47 &       0.00 &       7.47 \\
         2 &       0.86 &       6.47 &       0.61 &       6.62 \\
         3 &       1.48 &       5.90 &       1.10 &       6.08 \\
         4 &       2.25 &       5.36 &       1.86 &       5.50 \\
         5 &       2.99 &       4.93 &       2.59 &       5.05 \\
         6 &       3.73 &       4.59 &       3.34 &       4.62 \\
         7 &       4.66 &       4.26 &       4.10 &       4.21 \\
         8 &       6.17 &       3.72 &            &            \\
\hline
\end{tabular}  
\end{table}

\begin{table}
\center
\caption{Distance to the galactic plane ($z^{*}$ in kpc) and logarithmic 
space density, $D^{*}=\log D+10$, per unit absolute magnitude interval 
for stars with $5<M_{g}\leq10$ for the field SA 114 ($l=68.^{o}38$, 
$b=-48^{o}.38$).}
\begin{tabular}{ccc}
\hline
ID & $z^{*}$ & $D^{*}$ \\
\hline
1  & 0.00 & 7.52 \\
2  & 0.41 & 6.90 \\
3  & 0.63 & 6.62 \\
4  & 0.93 & 6.24 \\
5  & 1.30 & 5.91 \\
6  & 1.68 & 5.62 \\
7  & 2.06 & 5.42 \\
8  & 2.60 & 5.10 \\
9  & 3.36 & 4.71 \\
10 & 4.66 & 4.29 \\
11 & 6.46 & 3.81 \\
12 & 8.28 & 3.37 \\
13 & 10.21& 3.11 \\
\hline
\end{tabular}  
\end{table}

\begin{table*}
\setlength{\tabcolsep}{2pt} 
\center
\caption{Distance-to the galactic plane- dependant $a_{i}$ ($i$=0, 1, 2) parameters 
for different samples of absolute magnitude intervals (5,6] and (6,7] for the field $\#$0952+5245.}
\begin{tabular}{ccccccccccc}
\hline
$M_{g}\rightarrow$ & \multicolumn{4}{c}{(5,6]} &  & $M_{g}\rightarrow$ &   \multicolumn{4}{c}{(6,7]} \\
\hline
Sample No & $<z^{*}>$ & $z^{*}$ &     $D^{*}$ &                         &   &Sample No & $<z^{*}>$ & $z^{*}$& $D^{*}$ &    \\
\hline
 1        & 1.53      & 0.00 & 7.47 & $a_{2}=-0.3705$(0.0034) &   & 1        & 1.19      & 0.00 & 7.47 & $a_{2}=-0.6180$(0.0026)\\
          &           & 0.86 & 6.47 & $a_{1}=2.9923$(0.0077)  &   &          &           & 0.61 & 6.62 & $a_{1}=3.5886$(0.0050)\\
          &           & 1.48 & 5.90 & $a_{0}=5.8262$(0.0034)  &   &          &           & 1.10 & 6.08 & $a_{0}=5.8251$(0.0020)\\
          &           & 2.25 & 5.36 &                         &   &          &           & 1.86 & 5.50 &            \\
\hline
 2        & 2.24      & 0.00 & 7.47 & $a_{2}=-0.3095$(0.0305) &   & 2        & 1.85      & 0.00 & 7.47 & $a_{2}=-0.4710$(0.0761)\\
          &           & 1.48 & 5.90 & $a_{1}=2.8734$(0.0922)  &   &          &           & 1.10 & 6.08 & $a_{1}=3.3561$(0.2026)\\
          &           & 2.25 & 5.36 & $a_{0}=5.8315$(0.0601)  &   &          &           & 1.86 & 5.50 & $a_{0}=5.8338$(0.1136)\\
          &           & 2.99 & 4.93 &                         &   &          &           & 2.59 & 5.05 &            \\
\hline
 3        & 2.99      & 0.00 & 7.47 & $a_{2}=-0.2546$(0.0105) &   & 3        & 2.58      & 0.00 & 7.47 & $a_{2}=-0.3040$(0.0453)\\
          &           & 2.25 & 5.36 & $a_{1}=2.7246$(0.0384)  &   &          &           & 1.86 & 5.50 & $a_{1}=2.2685$(0.1502)\\
          &           & 2.99 & 4.93 & $a_{0}=5.8271$(0.0296)  &   &          &           & 2.59 & 5.05 & $a_{0}=5.8329$(0.1076)\\
          &           & 3.73 & 4.59 &                         &   &          &           & 3.34 & 4.62 &            \\
\hline
 4        & 3.79      & 0.00 & 7.47 & $a_{2}=-0.2183$(0.0099) &   & 4        & 3.34      & 0.00 & 7.47 & $a_{2}=-0.2062$(0.0222)\\
          &           & 2.99 & 4.93 & $a_{1}=2.6006$(0.0447)  &   &          &           & 2.59 & 5.05 & $a_{1}=2.6699$(0.0886)\\
          &           & 3.73 & 4.59 & $a_{0}=5.8272$(0.0419)  &   &          &           & 3.34 & 4.62 & $a_{0}=5.8286$(0.0726)\\
          &           & 4.66 & 4.26 &                         &   &          &           & 4.10 & 4.21 &            \\
\hline
 5        &  4.85     & 0.00 & 7.47 & $a_{2}=-0.1461$(0.0213) &   &          &           &       &     &            \\
          &           & 3.73 & 4.59 & $a_{1}=2.2933$(0.1288)  &   &          &           &       &     &            \\
          &           & 4.66 & 4.26 & $a_{0}=5.8336$(0.1704)  &   &          &           &       &     &            \\
          &           & 6.17 & 3.72 &                         &   &          &           &       &     &            \\
\hline
\end{tabular}  
\end{table*}

\begin{table*}
\setlength{\tabcolsep}{2pt} 
\center
\caption{Distance-to the galactic plane- dependant $a_{i}$ ($i$=0, 1, 2) parameters 
for different samples of absolute magnitude interval (5,10] for the field SA 114.}
\begin{tabular}{ccccccccccc}
\hline
Sample No & $<z^{*}>$ & $z^{*}$ &     $D^{*}$ &                         &   &Sample No & $<z^{*}>$ & $z^{*}$& $D^{*}$ &    \\
\hline
 1        & 0.99      & 0.00 & 7.52 & $a_{2}=-0.6939$(0.0522) &   & 5        & 2.87      & 0.00 & 7.52 & $a_{2}=-0.2895$(0.0302)\\
          &           & 0.41 & 6.90 & $a_{1}=3.7701$(0.0928)  &   &          &           & 1.68 & 5.62 & $a_{1}=2.9160$(0.1492)\\
          &           & 0.63 & 6.62 & $a_{0}=5.7082$(0.0344)  &   &          &           & 2.06 & 5.42 & $a_{0}=5.7740$(0.1725)\\
          &           & 0.93 & 6.24 &                         &   &          &           & 2.60 & 5.10 &            \\
          &           & 1.30 & 5.91 &                         &   &          &           & 3.36 & 4.71 &            \\
          &           & 1.68 & 5.62 &                         &   &          &           & 4.66 & 4.29 &            \\
\hline
 2        & 1.32      & 0.00 & 7.52 & $a_{2}=-0.6832$(0.0342) &   & 6        & 3.83      & 0.00 & 7.52 & $a_{2}=-0.2012$(0.0252)\\
          &           & 0.63 & 6.62 & $a_{1}=3.7569$(0.0742)  &   &          &           & 2.06 & 5.42 & $a_{1}=2.5807$(0.1739)\\
          &           & 0.93 & 6.24 & $a_{0}=5.7072$(0.0354)  &   &          &           & 2.60 & 5.10 & $a_{0}=5.8300$(0.2685)\\
          &           & 1.30 & 5.91 &                         &   &          &           & 3.36 & 4.71 &            \\
          &           & 1.68 & 5.62 &                         &   &          &           & 4.66 & 4.29 &            \\
          &           & 2.06 & 5.42 &                         &   &          &           & 6.46 & 3.81 &            \\
\hline
 3        & 1.71      & 0.00 & 7.52 & $a_{2}=-0.5462$(0.0633) &   & 7        & 5.07      & 0.00 & 7.52 & $a_{2}=-0.1428$(0.0249)\\
          &           & 0.93 & 6.24 & $a_{1}=3.5190$(0.1720)  &   &          &           & 2.60 & 5.10 & $a_{1}=2.2799$(0.2196)\\
          &           & 1.30 & 5.91 & $a_{0}=5.7495$(0.1085)  &   &          &           & 3.36 & 4.71 & $a_{0}=5.9073$(0.4266)\\
          &           & 1.68 & 5.62 &                         &   &          &           & 4.66 & 4.29 &            \\
          &           & 2.06 & 5.42 &                         &   &          &           & 6.46 & 3.81 &            \\
          &           & 2.60 & 5.10 &                         &   &          &           & 8.28 & 3.37 &            \\
\hline
 4        & 2.20      & 0.00 & 7.52 & $a_{2}=-0.3891$(0.0536) &   & 8        & 6.59      & 0.00 & 7.52 & $a_{2}=-0.1082$(0.0176)\\
          &           & 1.30 & 5.91 & $a_{1}=3.1897$(0.1889)  &   &          &           & 3.36 & 4.71 & $a_{1}=2.0541$(0.1901)\\
          &           & 1.68 & 5.62 & $a_{0}=5.7616$(0.1586)  &   &          &           & 4.66 & 4.29 & $a_{0}=5.9122$(0.4583)\\
          &           & 2.06 & 5.42 &                         &   &          &           & 6.46 & 3.81 &            \\
          &           & 2.60 & 5.10 &                         &   &          &           & 8.28 & 3.37 &            \\
          &           & 3.36 & 4.71 &                         &   &          &           &10.21 & 3.11 &            \\
\hline
\end{tabular}  
\end{table*}

\subsection{Estimation of the parameters for the absolute magnitude interval (5,6]}
We separated the eight ($z^{*}$, $D^{*}$) couples for the absolute magnitude 
interval (5,6] in Table 1 into five samples, i.e. (1,2,3,4); (1,3,4,5); (1,4,5,6); 
(1,5,6,7) and (1,6,7,8), each of which involves the couple (0,7.47) and estimated 
the $a_{i}$ ($i$=0,1,2) parameters in Eq. (10) by the least square method (Table 3).     
Since $a_{0}$ corresponds to the local space density of the stars with absolute 
magnitude (5,6], its value must be constant. Hence, we re-produced it for each 
sample and we found that the numerical value of the parameter $a_{0}$ is the 
same for each sample and equals to the Hipparcos' local space density for the 
absolute magnitude interval (5,6]. This result confirms suitability of our 
procedure we followed for the estimation of the $a_{i}$ parameters.

However, the trends of $a_{1}$ and $a_{2}$ are different than $a_{0}$ and from each other. 
For instance, $a_{1}$ decreases with increasing $z^{*}$ to the galactic plane, whereas $a_{2}$ 
increases in the same direction. This result also confirms the argument that the 
scaleheight of a population can not be adopted as a constant, but it should be 
(distance to galactic plane) dependent \citep{K06}. If $a_{2}$ is omitted 
in Eq. (10), the reciprocal of $a_{1}$ corresponds to scaleheight in Eq. (8) and 
increases with distance to galactic plane. 

\subsection{Estimation of the parameters for the absolute magnitude interval (6,7]}

The seven ($z^{*}$, $D^{*}$) couples for the absolute magnitude interval (6,7] in 
Table 1 were separated into four samples and the corresponding $a_{i}$ ($i$=0,1,2) 
parameters were estimated by the same procedure as was done and explained for 
the absolute magnitude interval (5,6]. Also, these results are given in 
Table 3. There, $a_{0}$ is found to be constant and equal the Hipparcos local 
space density, for the same absolute magnitude interval (6,7], $a_{0}$ = 7.47. 
On the other hand, the trends of $a_{1}$ and $a_{2}$ are the same as the 
corresponding ones for the absolute magnitude interval (5,6].

\subsection{Estimation of the parameters for stars with absolute magnitudes (5, 10]}

The number of ($z^{*}$, $D^{*}$) couples in Table 2 are 13 in total and 
more than in Table 1. We used this advantage to increase the sample 
numbers and the couple numbers in each sample. Thus, we separated the 
($z^{*}$, $D^{*}$) couples into eight samples, i.e. (1, 2, 3, 4, 5, 6); (1, 3, 4, 
5, 6, 7); (1, 4, 5, 6, 7, 8); (1, 5, 6, 7, 8, 9); (1, 6, 7, 8, 9, 10); (1, 7, 8, 
9, 10, 11); (1, 8, 9, 10, 11, 12) and (1, 9, 10, 11, 12, 13), each of which involves 
the couple (0,7.52) and estimated the $a_{i}$ ($i$=0,1,2) parameters in Eq. (10) by 
the procedure explained in section 4.1. The results are given in Table 4. 

Although the trends of $a_{2}$ and $a_{1}$ are the same as the trends of $a_{2}$ and $a_{1}$ 
for the data given in the previous sections, the value of $a_{0}$ is not constant, 
but it corresponds to a local space density within a range $7.43 \leq D^{*}(0)\leq 7.52$. 
The lower values correspond to the local space densities of brighter stars in the Hipparcos' 
catalogue. That is, the estimated local space density deviates from the mean local space density 
($<D^{*}>=7.52$) of stars with (5,10] as one goes to large vertical distances. This discrepancy 
is due to a bias effect, i.e. the galactic model parameters varies with distance 
\citep{Biliretal06, Aketal07, Cabreraetal07, K07, Biliretal08}.   

\section{Testing the unified density law}

We tested the unified density law by the ($z^{*}$, $D^{*}$) data for the absolute 
magnitude intervals (5,6], (6,7] and (5,10] for the ELAIS field making use of the 
following procedure. A mean $<z^{*}>$ distance were attributed to each sample in 
Table 3 and Table 4. Then, $a_{i}$ ($i$=0,1,2) parameters were evaluated for the given 
$z^{*}$ distance by interpolation/extrapolation of the corresponding $a_{i}$ ($i$=0,1,2) 
in Table 3 and Table 4. Finally, the space densities were evaluated according to Eq. (10). 
The results are presented in Table 5, Table 6 and Table7, for the absolute magnitude 
intervals (5,6], (6,7] and (5,10], respectively. For clarification, we state that the 
numbers in the sixth column are the sample numbers in Table 3 or Table 4, the $z^{*}$ 
distances in seventh column are the corresponding mean $z^{*}$ distances for these samples, 
$a_{i}$ ($i$=0,1,2) are the interpolated/extrapolated parameters for the $z^{*}$ on the 
same line, $D^{*}$ is the original logarithmic space density taken from \citet[][hereafter, 
BKG]{BKG06}, and $D^{*}_{ev}$ is the logarithmic space density evaluated by the interpolated/
extrapolated $a_{i}$ ($i$=0,1,2) parameters. The $\Delta D^{*}$ offsets of the evaluated 
logarithmic space densities from the original ones in the last columns in three tables 
mentioned above are rather small, confirming that the procedure could be applied efficiently.

We reproduced the $D^{*}$ logarithmic space densities for the z* distances given in 
Tables 5-7, for stars with absolute magnitudes (5,6], (6,7] and (5,10], making use of the 
calibrations of \cite {P00} and \cite {J08}, and we compared the $\Delta D^{*}$ offsets of 
the evaluated logarithmic space densities from the original ones with the corresponding 
offsets obtained by means of the unified density law. It is interesting to compare the 
results of \cite {P00} with the results we obtained. Since they estimated galactic model 
parameters in the vertical direction of the Galaxy, as was done in the present study. 
The galactic model parameters of \cite {P00} are as follows: Normalized local space 
densities for thin and thick discs $n_{1}=2.725\times10^{-3}$, $n_{2}=2.229\times
10^{-4}$, scaleheights for thin and thick discs $H_{1}$=280 pc, $H_{2}$=1267 pc, respectively. 
\cite {J08} estimated the following galactic model parameters both in vertical and 
longitudinal directions and they found for normalized local space densities $n_{1}=2.951\times
10^{-3}$, $n_{2}=3.160\times10^{-4}$, $n_{2}=1.317\times10^{-6}$, $H_{1}$=300 pc, 
$H_{2}$=900 pc, $(c/a)$=0.64. The evaluated logarithmic space densities, $D^{*}_{ev}$, 
and their offsets from the original ones taken from BKG are given in Tables 8, 9 and 10. 
The offsets were plotted in Fig. 1 and compared with the ones resulted by using the 
unified density law. One can see that there is a systematic deviation in the dispersions 
of the offsets evaluated via the calibrations of \cite {P00} and \cite {J08}, favouring 
the unified density law.

\begin{table}
\setlength{\tabcolsep}{1.8pt}
{\scriptsize 
\center
\caption{Space densities evaluated by the unified density law for stars 
with absolute magnitudes (5,6] in the ELAIS field. The columns 
give: (1) $z^{*}$ distance from the galactic plane in kpc, (2) the logarithmic 
space density D* taken from BKG, (3), (4) and (5): $a_{2}$, $a_{1}$ and $a_{0}$ 
parameters interpolated for the  $z^{*}$ distance on the same line, (6) no 
of samples used for interpolation of $a_{i}$ ($i$=1, 2, 3), (7) mean  $z^{*}$ 
distances for the samples used for the interpolation, (8) evaluated logarithmic 
space density $D^{*}_{ev}$, and (9) the difference between the evaluated 
logarithmic space density and the adopted one from BKG.}
\begin{tabular}{ccccccccc}
\hline
 $z^{*}$ &  $D^{*}$ &    $a_{2}$ &    $a_{1}$ &    $a_{0}$ & Sample no&  $<z^{*}>$ & $D^{*}_{ev}$ & $\Delta D^{*}$ \\
\hline
      1.60 &       5.66 &    -0.3645 &     2.9800 &     5.8267 &       1, 2 & 1.53 - 2.24 &       5.80 &      -0.14 \\
      1.93 &       5.51 &    -0.3361 &     2.9250 &     5.8292 &       1, 2 & 1.53 - 2.24 &       5.56 &      -0.05 \\
      2.46 &       5.17 &    -0.2934 &     2.8290 &     5.8302 &       2, 3 & 2.24 - 2.99 &       5.22 &      -0.05 \\
      3.15 &       4.77 &    -0.2456 &     2.6938 &     5.8271 &       3, 4 & 2.99 - 3.79 &       4.84 &      -0.07 \\
      4.35 &       4.17 &    -0.1804 &     2.4390 &     5.8305 &       4, 5 & 3.79 - 4.85 &       4.34 &      -0.17 \\
      6.15 &       3.80 &    -0.1461 &     2.2930 &     5.8336 &          5 &       4.85  &       3.74 &      +0.06 \\
\hline
\end{tabular}  
}
\end{table}

\begin{table}
\setlength{\tabcolsep}{1.8pt}
{\scriptsize 
\center
\caption{Space densities evaluated by the unified density law for stars with 
absolute magnitudes (6,7] in the ELAIS field. The symbols are as in Table 5.}
\begin{tabular}{ccccccccc}
\hline
$z^{*}$ &    $D^{*}$ &    $a_{2}$ &    $a_{1}$ &    $a_{0}$ &  Sample no &  $<z^{*}>$ & $D^{*}_{ev}$ & $\Delta D^{*}$ \\
\hline
      1.24 &       5.94 &    -0.6095 &     3.5710 &     5.8258 &       1, 2 & 1.19 - 1.85 &       5.95 &      -0.01 \\
      1.58 &       5.63 &    -0.5323 &     3.4512 &     5.8302 &       1, 2 & 1.19 - 1.85 &       5.68 &      -0.05 \\
      1.94 &       5.31 &    -0.4482 &     3.3094 &     5.8337 &       2, 3 & 1.85 - 2.58 &       5.41 &      -0.10 \\
      2.45 &       5.11 &    -0.3360 &     3.0448 &     5.8331 &       2, 3 & 1.85 - 2.58 &       5.10 &      +0.01 \\
      3.18 &       4.73 &    -0.2234 &     2.7351 &     5.8295 &       3, 4 & 2.58 - 3.34 &       4.67 &      +0.06 \\
\hline
\end{tabular}  
}
\end{table}

\begin{table}
\setlength{\tabcolsep}{1.8pt}
{\scriptsize 
\center
\caption{Space densities evaluated by the unified density law for thin disc, 
thick disc and halo components of the Galaxy for stars with absolute magnitudes 
(5,10] in the ELAIS field. The symbols are as in Table 5.}
\begin{tabular}{ccccccccc}
\hline
$z^{*}$ &    $D^{*}$ &    $a_{2}$ &    $a_{1}$ &    $a_{0}$ &  Sample no &  $<z^{*}>$ & $D^{*}_{ev}$ & $\Delta D^{*}$ \\
\hline
0.92  & 6.20 &  -0.6961 &  3.7729 & 5.7085 &  1, 2 &  $<$0.99     & 6.27 & -0.07 \\ 
1.26  & 5.86 &  -0.6851 &  3.7593 & 5.7074 &  1, 2 &  0.99 - 1.32 & 5.94 & -0.08 \\
1.61  & 5.55 &  -0.5821 &  3.5813 & 5.7384 &  2, 3 &  1.32 - 1.71 & 5.66 & -0.11 \\
1.96  & 5.35 &  -0.4697 &  3.3520 & 5.7557 &  3, 4 &  1.71 - 2.20 & 5.43 & -0.08 \\
2.52  & 4.95 &  -0.3416 &  3.0592 & 5.7675 &  4, 5 &  2.20 - 2.87 & 5.09 & -0.14 \\
3.22  & 4.59 &  -0.2573 &  2.7936 & 5.7944 &  5, 6 &  2.87 - 3.83 & 4.73 & -0.14 \\ 
4.58  & 4.03 &  -0.1658 &  2.3986 & 5.8768 &  6, 7 &  3.83 - 5.07 & 4.19 & -0.16 \\
6.03  & 3.56 &  -0.1210 &  2.1377 & 5.9104 &  7, 8 &  5.07 - 6.59 & 3.67 & -0.11 \\
8.04  & 3.15 &  -0.0753 &  1.8395 & 5.9168 &  7, 8 &  $>$6.59     & 3.12 &  0.03 \\
\hline
\end{tabular}  
}
\end{table}

\begin{table}
\center
\caption{Comparison of the original logarithmic space densities taken from BKG 
$D^{*}$ and the evaluated ones $D^{*}_{ev}$ by means of the Galactic model 
parameters of \citet{P00} (columns 3 and 4) and \citet{J08} (columns 5 and 6) 
for the absolute magnitude interval (5,6].}
\begin{tabular}{cccccc}
\hline
 $z^{*}$ &  $D^{*}$ & $D^{*}_{ev}$ & $\Delta D^{*}$ & $D^{*}_{ev}$ & $\Delta D^{*}$ \\
\hline
      1.60 &       5.66 &       5.87 &      -0.21 &       5.83 &      -0.17 \\
      1.93 &       5.51 &       5.73 &      -0.22 &       5.62 &      -0.11 \\
      2.46 &       5.17 &       5.54 &      -0.37 &       5.33 &      -0.16 \\
      3.15 &       4.77 &       5.31 &      -0.54 &       4.98 &      -0.21 \\
      4.35 &       4.17 &       4.92 &      -0.75 &       4.40 &      -0.23 \\
      6.15 &       3.80 &       4.34 &      -0.54 &       3.60 &      +0.20 \\
\hline
\end{tabular}  
\end{table}

\begin{table}
\center
\caption{Comparison of the original logarithmic space densities taken from BKG 
($D^{*}$) and the evaluated ones ($D^{*}_{ev}$) by means of the galactic model 
parameters of \citet{P00} (columns 3 and 4) and \citet{J08} (columns 5 and 6) 
for the absolute magnitude interval (6,7].}
\begin{tabular}{cccccc}
\hline
        $z^{*}$ &   $D^{*}$ & $D^{*}_{ev}$ & $\Delta D^{*}$ & $D^{*}_{ev}$ & $\Delta D^{*}$ \\
\hline
      1.24 &       5.94 &       6.08 &      -0.14 &       6.09 &      -0.15 \\
      1.58 &       5.63 &       5.88 &      -0.25 &       5.84 &      -0.21 \\
      1.94 &       5.31 &       5.73 &      -0.42 &       5.62 &      -0.31 \\
      2.45 &       5.11 &       5.54 &      -0.43 &       5.33 &      -0.22 \\
      3.18 &       4.73 &       5.30 &      -0.57 &       4.93 &      -0.20 \\
\hline
\end{tabular}  
\end{table}

\begin{table}
\center
\caption{Comparison of the original logarithmic space densities taken from BKG 
($D^{*}$) and the evaluated ones ($D^{*}_{ev}$) by means of the galactic model 
parameters of \citet{P00} (columns 3 and 4) and \citet{J08} (columns 5 and 6) 
for the absolute magnitude interval (5,10].}
\begin{tabular}{cccccc}
\hline
        $z^{*}$ &   $D^{*}$ & $D^{*}_{ev}$ & $\Delta D^{*}$ & $D^{*}_{ev}$ & $\Delta D^{*}$ \\
\hline
0.92 & 6.20 & 6.32 & -0.12 & 6.41 & -0.21\\
1.26 & 5.86 & 6.05 & -0.19 & 6.09 & -0.23\\
1.61 & 5.55 & 5.85 & -0.30 & 5.83 & -0.28\\
1.96 & 5.35 & 5.70 & -0.35 & 5.61 & -0.26\\
2.52 & 4.95 & 5.49 & -0.54 & 5.30 & -0.35\\
3.22 & 4.59 & 5.25 & -0.66 & 4.94 & -0.35\\
4.58 & 4.03 & 4.78 & -0.75 & 4.28 & -0.25\\
6.30 & 3.56 & 4.19 & -0.63 & 3.55 & +0.01\\
8.04 & 3.15 & 3.59 & -0.44 & 3.02 & +0.13\\
\hline
\end{tabular}  
\end{table}

\begin{figure}
\begin{center}
\includegraphics[scale=0.38, angle=0]{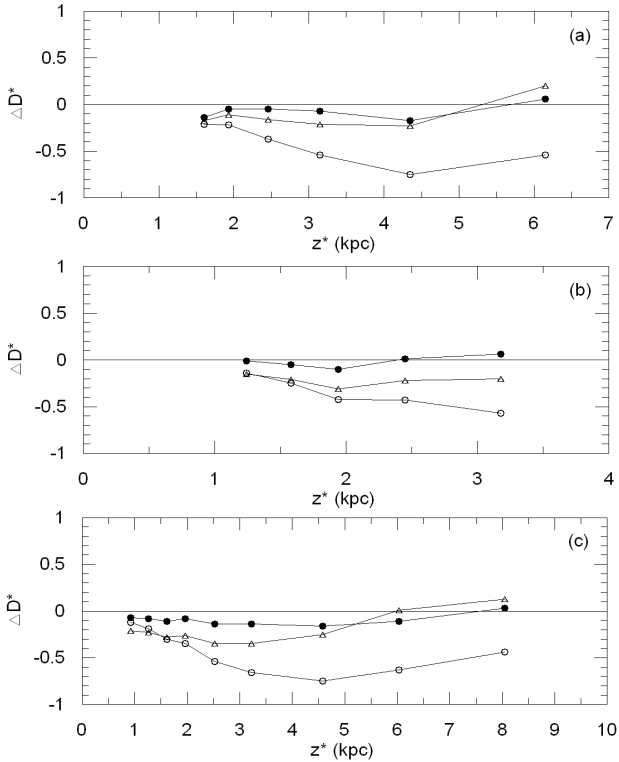}
\caption[]{Offsets of the evaluated logarithmic space densities ($\Delta D^{*}$) 
versus distance from the galactic plane ($z^{*}$) for the absolute magnitude 
intervals (5,6] (panel a), (6,7] (panel b) and (5,10] (panel c). Symbols:  
This study ($\bullet$), \citet{P00} ($\circ$), and \citet{J08} ($\bigtriangleup $).}
\end{center}
\end{figure}

 \section{Discussion} 
We used a unified density law for the thin disc, thick disc and halo 
to match the observational logarithmic space densities evaluated for the 
fields $\#$0952+5245 and SA 114 in the vertical direction up to 8 kpc 
from the galactic plane. The exponent of the density law is adopted as a quadratic 
function of the distance from the galactic plane. The absolute magnitude dependent 
parameters estimated for a set of distance from the galactic plane, were 
interpolated/extrapolated for three given sets of $z$--distances 
and applied to the ELAIS field. The offsets of the logarithmic space densities 
evaluated by the interpolated/extrapolated parameters from the original 
values are rather small. Whereas the offsets corresponding to the 
values of galactic model parameters of \citet{P00} and \citet{J08} are 
considerable systematic and large, favouring our density law.

The so called ``unified density law'' does not define any population type, up to 
distances of 8 kpc from the galactic plane. Hence, the space density of the Galaxy 
could be matched to a unified density law with three parameters. This approach would  
imply and support the argument of existence of a single disc. Hence, the thin and 
thick discs discussed for a long time in the literature are not discrete. However, 
we reiterate that this is not our idea. But we confirmed their arguments 
\citep[cf.][]{N87, I08}. A significant fraction of material with $[Fe/H]<-1$ dex has 
disclike structure \citep{N87}. This confirms our argument related to the 
unified density law. 

In the present work, we proposed galactic model parameters as a function of distance 
from the galactic plane, as was done in the study of \citet{K06}. However, there is 
a difference between, what was proposed over there and here. Namely, in the former 
study, scaleheight and scalelength were taken into consideration, whereas in the present 
study, none of these parameters are taken into consideration. If we omit the 
coefficient of $z^{2}$, for instance $a_{2}$, reciprocal of the coefficient of $z$, 
$1/a_{1}$ corresponds to scaleheight for the total density, but not for a specific 
population.

\section{Conclusion} 
\citet{N87} proposed a Galaxy model where he did not assume thin and thick discs 
are discrete components. Rather, he assumed that the thin and thick discs form a 
kinematical and chemical continuum. \citet{I08} confirmed this hypothesis by 
demonstrating the absence of a correlation between the observed velocity and 
metallicity distributions for disc stars. The present work, reinforces the same 
argument, making use of different procedure. The density law proposed here can 
match to the observed vertical space densities up to several kpc without the need 
to separate the Galactic stars into separate population types.

\section{Acknowledgments}
S. Karaali  and E. Hamzao\u glu are grateful to the Beykent University for financial 
support.


\end{document}